\documentclass[twocolumn,prl,showpacs]{revtex4}
\usepackage{graphicx,color}
\usepackage{dcolumn}
\usepackage{amsmath,amssymb}

\def\wJA{1/0.635}
\def\wTRI{5.38}
\def\zTRI{5.03}
\def\cTRI{0.859}
\def\xonetwoTRI{0.293}
\def\xonethreeTRI{1.439}
\def\rmTRI{25\%}

\newcommand{\myphi}[3]{\phi_{#1,#2}^{(#3)}}
\newcommand{\myx  }[3]{   x_{#1,#2}^{(#3)}}

\def\figurescalelargeA{0.95}
\def\figurescalelargeB{0.90}
\def\figurescalelargeC{0.85}

 \def\ZUichi{%
 \begin{figure}[t]
  \begin{center}
   \includegraphics[width=\figurescalelargeA\linewidth]{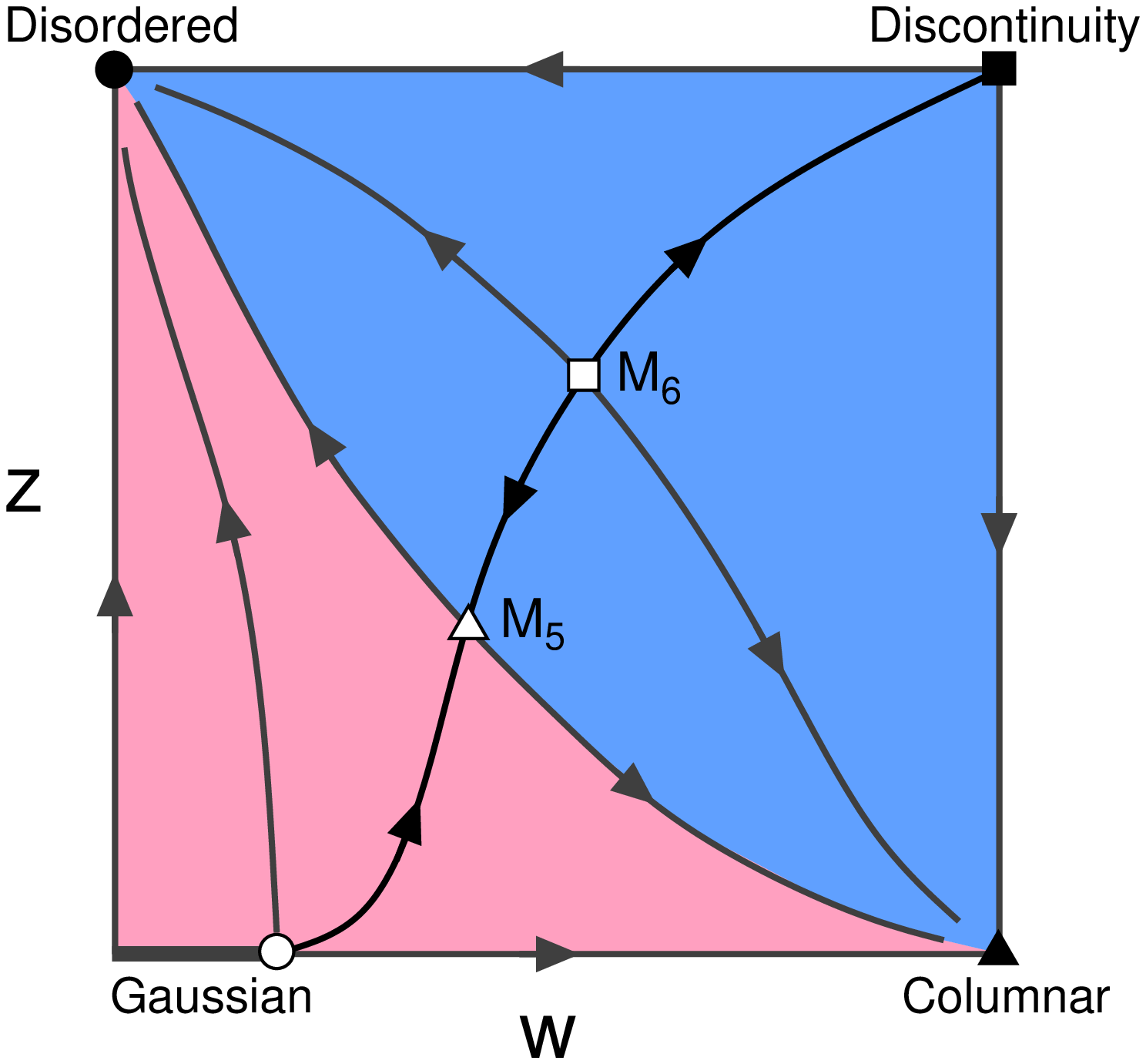}
  \end{center}
  \vspace{-5mm}
  \caption{%
  (Color online)
  A rough sketch of the RG flow diagram.
  The open circle denotes the end point of the Gaussian fixed line on
  the $w$ axis; the marks for fixed points are given with explanations.
  The sine-Gordon theory (the Potts lattice gas theory) is varied in the
  red (blue) region.
  }
  \label{RGflow}
 \end{figure}
 }

 \def\ZUni{%
 \begin{figure}[t]
  \begin{center}
   \includegraphics[width=\figurescalelargeC\linewidth]{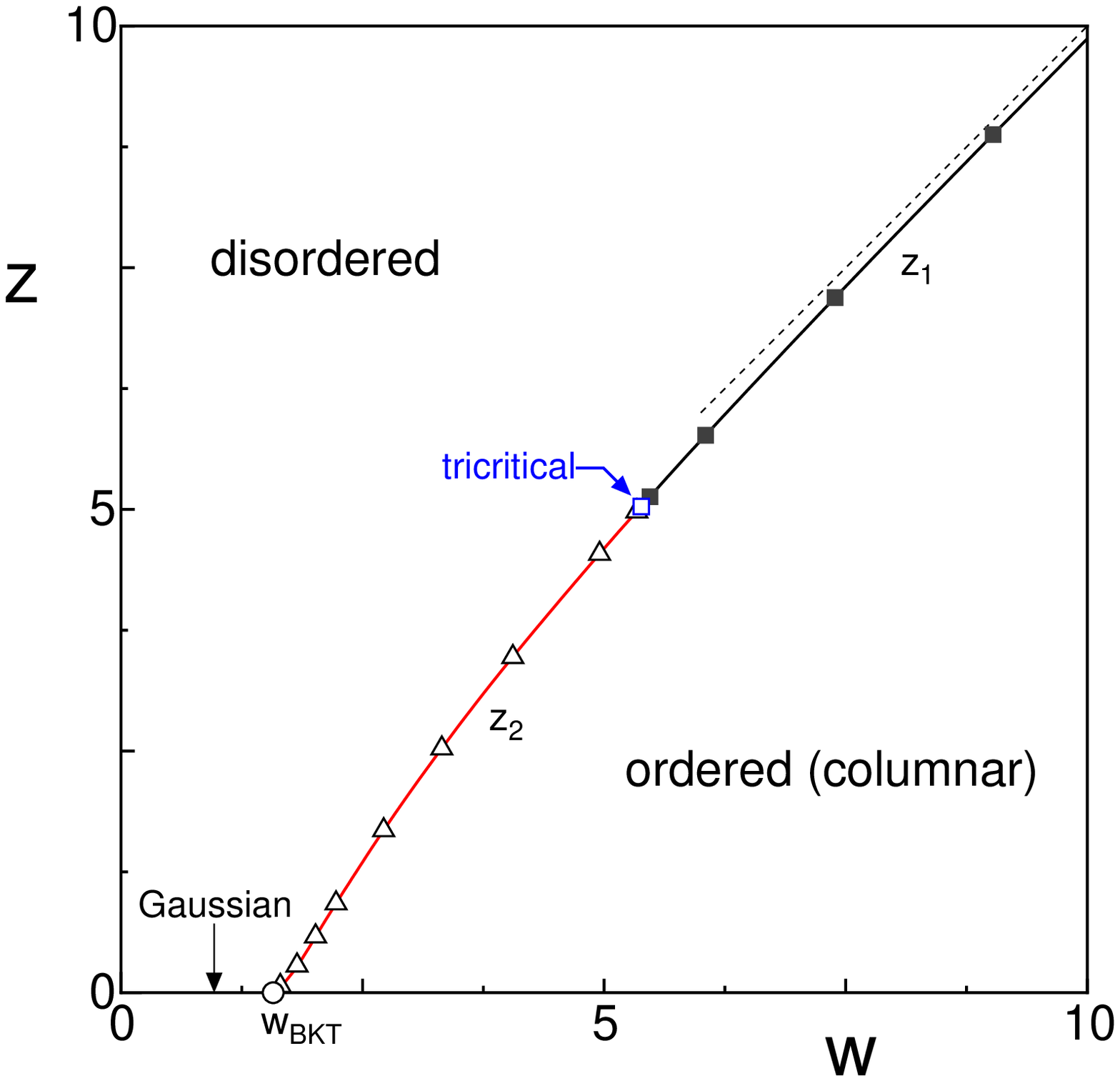}
  \end{center}
  \vspace{-5mm}
  \caption{%
  (Color online)
  The global phase diagram.
  Open triangles (filled squares) represent the second-order
  (first-order) phase transition points; the open square denotes the
  tricritical point.
  The dotted line gives the asymptotic form $z=w$.
  }
  \label{phase}
 \end{figure}
 }

 \def\ZUsan{%
 \begin{figure}[t]
  \begin{center}
   \includegraphics[width=\figurescalelargeB\linewidth]{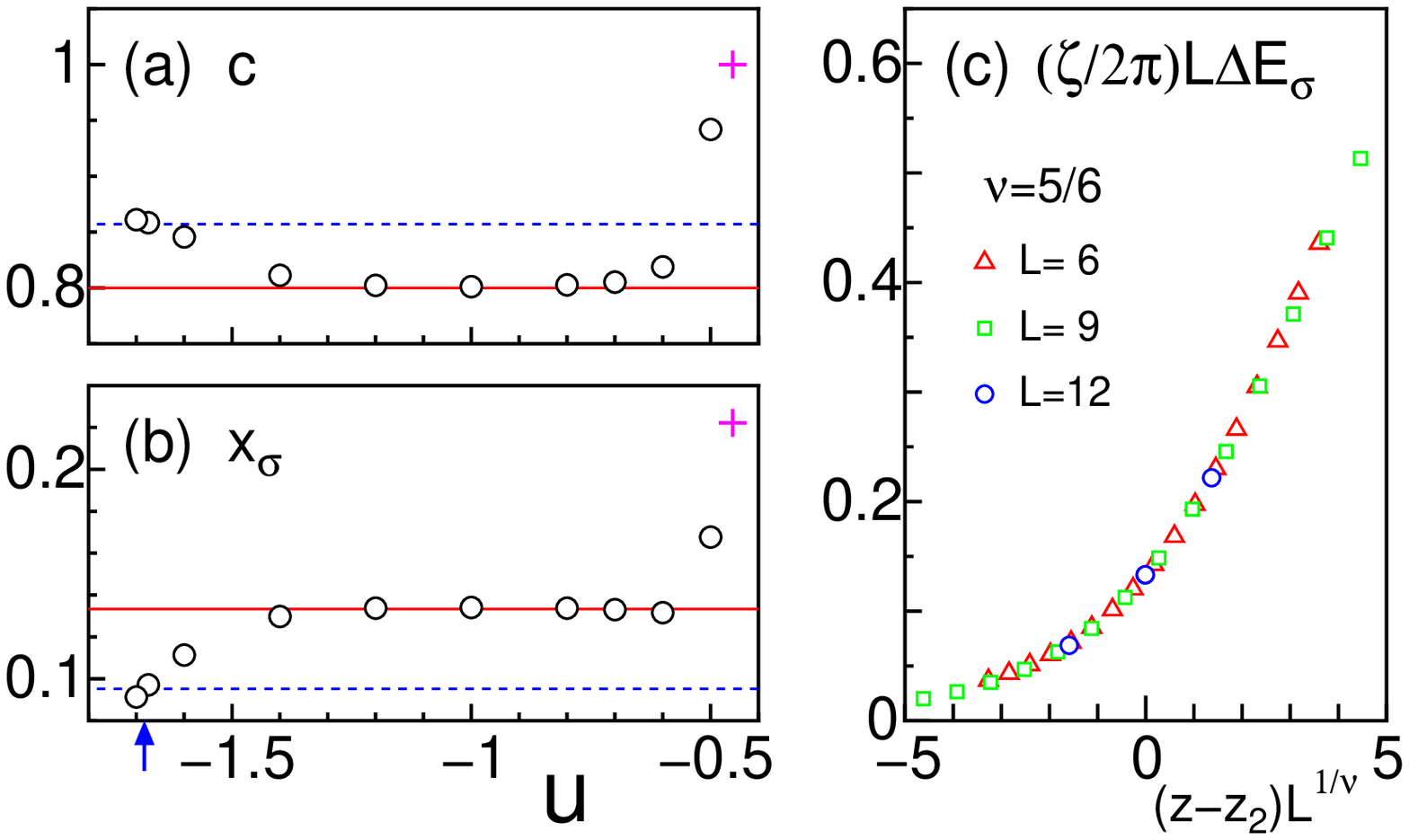}
  \end{center}
  \vspace{-5mm}
  \caption{%
  (Color online)
  The estimations (open circles) of (a) $c$ and (b) $x_\sigma$ along
  $z_2(w)$.
  Solid and dotted lines denote theoretical values for ${\cal M}_5$ and
  ${\cal M}_6$; the crosses also give those at the BKT point.
  The vertical arrow indicates the tricritical point.
  (c) The finite-size-scaling plot of $\Delta E_\sigma$ at $u=-1$.
  }
  \label{CFT}
 \end{figure}
 }

\begin{document}

 \title{%
 Monomer-Dimer Mixture on a Honeycomb Lattice}

 \author{%
 Hiromi Otsuka}

 \affiliation{%
 Department of Physics, Tokyo Metropolitan University, Tokyo 192-0397,
 Japan}
 
 \date{\today}

\begin{abstract}
 We study a monomer-dimer mixture defined on a honeycomb lattice as a
 toy model for the spin ice system in a magnetic field.
 In a low-doping region of monomers, the effective description of this
 system is given by the dual sine-Gordon model.
 In intermediate- and strong-doping regions, the Potts lattice gas
 theory can be employed.
 Synthesizing these results, we construct a renormalization-group flow
 diagram, which includes the stable and unstable fixed points
 corresponding to ${\cal M}_5$ and ${\cal M}_6$ in the minimal models of
 the conformal field theory.
 We perform numerical transfer-matrix calculations to determine a global
 phase diagram and also to proffer evidence to check our prediction.
\end{abstract}


 \maketitle

 An introductory study of dimer degrees of freedom for condensed matter
 physics was carried out long time ago.
 In a paper by Fowler and Rushbrooke in 1937, the dimer represented a
 diatomic molecule adsorbed on a crystal surface
 \cite{Fowl37}.
 Later, the statistical mechanics of dimers was studied by Kasteleyn
 \cite{Kast61}
 and
 Temperley and Fisher
 \cite{Temp61Fish61}.
 These pioneering works have provided exact solutions of two-dimensional
 models under certain conditions. 
 Since then,
 dimer models have gathered increasing attention.

 The correspondence between dimer models and real substances seems to
 fall into two categories.
 First, like for the case of diatomic molecules, the correspondence
 originates in the shape of the elements.
 For example,
 recently
 a network system made of rodlike molecules on a
 substrate was measured by using scanning tunneling microscopy
 \cite{Blun08}.
 It was argued that, because of the shape of the molecules, a
 honeycomb-lattice dimer model can be employed to reproduce properties
 of the network
 \cite{Jaco09}.
 Second, the correspondence can originate in interactions between
 elements.
 Spin configurations in the ground state of a triangular-lattice
 antiferromagnetic Ising model can be related to those of dimers on a
 honeycomb lattice
 \cite{Blot82}.
 In this case, dimer degrees of freedom emerge on a different lattice
 and represent unsatisfied bonds.

 In this work, we investigate a dimer-based model which can be related,
 in the latter sense, to the spin ice in a magnetic field.
 To describe their relevance, we begin by summarizing related
 researches.
 The rare-earth titanates such as $R_2$Ti$_2$O$_7$ ($R={\rm Ho}$, Dy)
 are known as the Ising pyrochlore magnets (IPM)
 \cite{Harr97Rami99},
 where each $R^{3+}$ behaves as an Ising spin along a local axis
 pointing
 in
 the center of a tetrahedron.
 In the low temperature, due to the magnetic couplings
 \cite{Hert00},
 six states, satisfying the so-called 2-in-2-out condition (also known
 as the ice rule), are permitted for each tetrahedron.
 When a magnetic field is applied along its [111] direction, a plateau
 is observed at $\frac23$ of the saturation magnetization
 \cite{Mats02Hiro03,Saka03}.
 This incompressible state is called the kagome ice.
 For a kagome layer sandwiched between two triangular layers, only
 three states are permitted for each tetrahedron pointing in the $[111]$
 ($[\bar 1\bar 1\bar 1$]) direction since spins on the triangular layers
 are polarized.
 Then, the mapping from the spin configuration on the kagome layer to
 the dimer configuration on the honeycomb lattice is possible, which
 enables us to exactly enumerate the residual entropy
 \cite{Moes01Udag02}.

 While the magnetic field causes a dimensional reduction, it also
 induces low-energy excitations that lead to breakings of the ice rule:
 For each tetrahedron pointing in the $[111]$ ($[\bar 1\bar 1\bar 1$])
 direction, a state that minimizes the Zeeman energy takes the
 3-in-1-out (1-in-3-out) configuration.
 This state can be viewed as a magnetic monopole (antimonopole) in the
 solid
 \cite{Cast08,Monopoles,Kado09,Isak04}.
 Specific heat measurement revealed that the temperature dependence of
 the monopole density obeys the Arrhenius law with an energy scale
 controlled by the magnetic field
 \cite{Kado09}.
 So, monopole excitation effects become significant at least around the
 phase boundary between the kagome ice and a saturated state
 \cite{Cast08,Isak04}.
 We note that since monopole-antimonopole pair creations and
 annihilations are only permitted, it follows naturally to introduce
 monomers in the dimer model to describe them.

 Motivated by the research advances on IPM, we consider a monomer-dimer
 mixture (MDM) on a honeycomb lattice $\Lambda_{\rm H}$.
 When we write the dimer occupation number on the kagome lattice
 $\Lambda_{\rm K}$ (the medial lattice of $\Lambda_{\rm H}$) as
 $n_l=0,1$, then a reduced Hamiltonian is given by
 \begin{equation}
  \beta H
   =
   u\!
   \sum_{\rho={\rm a,b,c}^{}}
   \sum_{\langle l,m\rangle\in\Lambda_{\rm K}^\rho}
   n_l n_m
   -
   \mu
   \sum_{l\in\Lambda_{\rm K}}
   n_l
   \label{Hamil}
 \end{equation}
 ($\Lambda_{\rm K}^{\rm a,b,c}$, three sublattices of $\Lambda_{\rm K}$).
 The first term represents the interaction between two neighboring
 dimers and the second term signifies the chemical potential that
 controls the density of dimers (and thus monomers).
 In addition, we impose the hard-core constraint,
 $\forall j\in\Lambda_{\rm H}$, $\sum_{l\in \{l(j)\}}n_l=0,1$,
 where $\{l(j)\}$ denotes three sites around $j$.
 For the compounds under consideration, the long-range dipole
 interaction is expected to be large
 \cite{Hert00}.
 While the so-called projective equivalence explains their spin-ice
 behaviors
 \cite{Isak05Taba06},
 the long-range nature may become important with breaking the ice rule.
 Here, we have employed the oversimplified model with the short-range
 interaction because we focus on universal properties stemming from a
 competition between the interaction and the monomer doping effects,
 which could shed some light on the understanding of, for instance, the
 properties of monopoles in the kagome ice.

 The MDM system on the square lattice was discussed by several authors
 \cite{Alet05,Papa07},
 where an effective theory for long-distance behaviors plays an
 important role.
 We take a similar approach; however our analysis predicts the emergence
 of criticalities not observed in the square lattice case.
 Let us start with the dimer covering case without the interaction
 ($u=0$, $\mu=\infty$).
 Its critical behaviors are described by a two-dimensional sine-Gordon
 Lagrangian density
 ${\cal L}_0={\cal L}_{\rm G}+{\cal L}_{\varphi}$
 with 
 \begin{align}
  {\cal L}_{\rm G}
  =
  \frac{K}{2\pi}
  \left(\nabla{\varphi}\right)^2,~~~
  {\cal L}_{\varphi}
  =
  \frac{y_\varphi}{2\pi\alpha^2}:\cos3\sqrt2\varphi:.
 \end{align}
 We have denoted the
 coarse-grained
 height field as $\varphi({\bf x})$,
 which satisfies the periodicity condition
 $\sqrt2\varphi=\sqrt2\varphi+2\pi N$ with $N\in\mathbb{Z}$
 ($\alpha$, a short distance cutoff)
 \cite{Alet05,Henl97}.
 Since the Gaussian coupling $K$ equals $\frac12$, the nonlinear term
 ${\cal L}_{\varphi}$ that represents the discreteness in the original
 height field is irrelevant
 ($y_\varphi$, a negative constant).

 Next, we consider a modification of ${\cal L}_0$ that is brought about
 by the nearest-neighbor interaction.
 Jacobsen and Alet discussed the same effect in a somewhat different
 context
 \cite{Jaco09}
 and concluded that, like the square-lattice case, it is devoted to a
 renormalization of the Gaussian coupling
 \cite{Papa07,Otsu09}.
 Thus, we can write $K(u)\simeq\frac12+c_1u$ ($c_1$, a negative
 constant).
 On the other hand, a monomer on, say, the A (B) sublattice of
 $\Lambda_{\rm H}$ corresponds to a defect with a positive (negative)
 charge.
 In the sine-Gordon language, using the disorder field $\vartheta$ dual
 to $\varphi$, we can express monomers as
 ${\rm e}^{\pm{\rm i}\sqrt2\vartheta}$.
 Consequently, at least in the low-doping region, a dual sine-Gordon
 model
 ${\cal L}={\cal L}_{\rm G}(u)+{\cal L}_{\varphi}+{\cal L}_{\vartheta}$
 describes the MDM system,
 where
 \begin{align}
  {\cal L}_{\vartheta}
  =
  \frac{y_\vartheta}{2\pi\alpha^2}:\cos\sqrt2\vartheta:
 \end{align}
 and a fugacity of monomers
 $y_\vartheta\propto{\rm e}^{-\mu/2}$.
 To proceed further, we clarify a renormalization-group (RG) flow in the
 attractive region $u\le0$.
 Since the dimensions of
 $\cos3\sqrt2\varphi$ and $\cos\sqrt2\vartheta$ are $9/2K$ and $K/2$,
 respectively, the Gaussian fixed line is realized for
 $\frac12\le K\le\frac94$ in the dimer covering case, whereas it is
 unstable against the doping.
 We sketch the flow in Fig.\ \ref{RGflow} (see the red region), where
 $(w,z)=({\rm e}^{-u},{\rm e}^{-\mu/2})$.
 The open circle on the $w$ axis denotes the end point $K=\frac94$,
 where ${\cal L}_{\varphi}$ becomes marginal and brings about the
 Berezinskii-Kosterlitz-Thouless (BKT) transition to the 3-fold
 degenerate columnar ordered state, as shown by the flow to the filled
 triangle.
 Meanwhile the relevant ${\cal L}_{\vartheta}$ leads to the disordered
 phase (see the flow to the filled circle).
 ${\cal L}_{\varphi}$ and ${\cal L}_{\vartheta}$ are mutually non-local,
 so intrinsically they are in competition. 
 If both of them are relevant, it becomes obvious as a flow to an
 infrared (IR) fixed point.
 Since this fixed point describes the transition to the 3-fold
 degenerate ordered state, it may be of the 3-state Potts universality
 class.
 Consequently, we can expect a massless RG flow from the ultraviolet
 (UV) fixed point with the central charge $c=1$ (the open circle) to the
 IR fixed point with $c=\frac45$ (the open triangle)
 \cite{Lech02}.
 Indeed, ${\cal L}$ is an effective theory for the 3-state Potts model,
 and, under the self-dual condition, this flow has been explicitly
 obtained and understood as a renormalization of the ${\mathbb Z}_3$
 neutral $X$ operator, with dimension $x_X=\frac{14}{5}$, to the IR
 fixed point
 \cite{Lech02}. 
 Although ${\cal L}$ is not self-dual around the open circle, we suppose
 that the same RG flow should be observed at least around the open
 triangle.
 We provide numerical evidence for this assumption below.

 \ZUichi

 For the exploration of intermediate- and strong-doping regions, let us
 focus on the role of the $X$ operator.
 Writing the minimal model series of the conformal field theory with
 $c=1-6/p(p+1)$ as ${\cal M}_p$ and a primary field at the position
 $(r,s)$ of the conformal grid as $\myphi{r}{s}{p}$, then
 $X=\myphi{3}{1}{5}$, i.e., the {\it leading irrelevant} operator on
 ${\cal M}_5$.
 Zamolodchikov
 \cite{Zamo87}
 and
 Ludwig and Cardy
 \cite{Ludw87}
 discussed the deformation of ${\cal M}_p$ by the {\it least relevant}
 operator $\myphi{1}{3}{p}$ and concluded that there exists a RG flow
 connecting minimal models as
 \begin{align}
  {\rm UV}: {\cal M}_{p  },~\myphi{1}{3}{p  }
  ~\longrightarrow~
  {\rm IR}: {\cal M}_{p-1},~\myphi{3}{1}{p-1},
  \label{UVIR1}
 \end{align}
 where a UV-IR operator transmutation is also given.
 By taking these into account, it is plausible that another UV fixed point
 exists and that a RG flow connects ${\cal M}_6$ (the open square) and
 ${\cal M}_5$ (the open triangle), as given in Fig.\ \ref{RGflow}.
 In the opposite direction, the renormalization always amplifies a
 deviation from ${\cal M}_6$, so the coupling of $\myphi{1}{3}{6}$ flows
 to a strong-coupling fixed point (the filled square).
 It may be a discontinuity one, and thus the transition is of first
 order in the strong-doping region.

 The emergence of the new critical fixed point ${\cal M}_6$ with
 $c=\frac67$ is key to understanding the MDM system.
 We take a look at the following UV-IR operator correspondence and
 characterize it
 \cite{Ludw87}.
 \begin{align}
  {\rm UV}: \myphi{1}{3}{6},~\myphi{5}{5}{6},~\myphi{3}{3}{6}
  ~\longrightarrow~
  {\rm IR}: \myphi{3}{1}{5},~\myphi{2}{1}{5},~\myphi{3}{3}{5}.
  \label{UVIR2}
 \end{align}
 The latter two of the three IR operators correspond to the energy
 density ($\varepsilon$) and the ${\mathbb Z}_3$ spin ($\sigma$) of the
 3-state Potts model, respectively.
 Thus,
 $\myphi{1}{2}{6}$ ($=\myphi{5}{5}{6}$) as well as
 $\myphi{1}{3}{6}$ is ${\mathbb Z}_3$ neutral, and represents a thermal
 operator in the Lagrangian.
 Since, using the Kac formula, dimensions of UV operators are given as
 $\myx{1}{3}{6}=\frac{10}{ 7}$,
 $\myx{1}{2}{6}=\frac{ 2}{ 7}$, and
 $\myx{3}{3}{6}=\frac{ 2}{21}$,
 $\myphi{1}{3}{6}$ and $\myphi{1}{2}{6}$ are both relevant and provide
 the flow around the open square in Fig.\ \ref{RGflow}.
 In the literature, one can find a model that exhibits the same RG flow
 given in the blue region of Fig.\ \ref{RGflow}.
 Nienhuis discussed the $q$-state Potts lattice gas model on a square
 lattice
 \cite{Nien82},
 where the leading and next-leading thermal exponents were obtained on
 both the critical and tricritical points.
 Indeed, for $q=3$, those on the latter are
 $y_1=\frac{4}{7}$ and
 $y_2=\frac{12}{7}$, which agree with the dimensions
 $\myx{1}{3}{6}$ and
 $\myx{1}{2}{6}$.
 Therefore, the MDM system in the intermediate- and strong-doping
 regions can be viewed as the 3-state Potts lattice gas, where monomers
 and dimers play roles of the vacancies and the ${\mathbb Z}_3$ spins,
 respectively.

 Obviously, the 3-fold axis symmetry in $\Lambda_{\rm H}$ plays a
 decisive role in determining ${\cal L}_\varphi$ and mapping to the
 Potts lattice gas with $q=3$.
 In contrast, for the MDM on the square lattice
 \cite{Alet05,Papa07},
 due to the 4-fold axis symmetry, the sine-Gordon theory with a
 potential $\cos4\sqrt2\varphi$ and a 4-state Potts lattice gas theory
 are relevant in the weak- and strong-doping regions, respectively.
 So, instead of crossovers of criticalities, we observe a fixed line
 that terminates by the first-order phase transition line.

 Now, we explain numerical calculations and results to check our
 prediction.
 We have focused on a structure of the phase diagram and the
 universality classes of phase transitions.
 We first summarize our results on the phase diagram in Fig.\
 \ref{phase}.
 We quote the BKT point of
 \cite{Jaco09},
 $w_{\rm BKT}\simeq\wJA$,
 which is indicated by the open circle on the $w$ axis.
 The curve for open triangles (filled squares) denotes the second-order
 (first-order) phase transition boundary $z_2(w)$ [$z_1(w)$] between the
 ordered and disordered phases. The open square represents the
 tricritical point.
 For their enumerations, we have performed numerical transfer-matrix
 calculations of Eq.\ (\ref{Hamil}) on $\Lambda_{\rm H}$ with
 $L\times\!\infty$ cylinder geometry [$L$, the circumference of the
 cylinder in units of the length between two parallel bonds, is given in
 multiples of 3 because the columnar state is expected].
 We note that, with the monomer doping, the row-to-row transfer matrix
 ${\bf T}(L)$ becomes less sparse.
 Thus, the accessible system size is strongly limited to a small number,
 e.g., $L\le12$ in our calculations.

 \ZUni

 To determine the second-order phase transition points, we have
 performed phenomenological RG (PRG) calculations.
 Writing the eigenvalues of ${\bf T}(L)$ as $\lambda_i(L)$ and their
 logarithms as $E_i(L)=-\ln|\lambda_i(L)|$ ($i$ specifies an
 excitation), then the conformal invariance provides the direct
 expressions of $c$ and $x_i$ in the critical systems as
 $E_{\rm g}(L)\simeq Lf-\pi c/6L\zeta$
 and
 $\Delta E_i(L)\simeq 2\pi x_i/L\zeta$
 \cite{Blot86,Card84}.
 $E_{\rm g}(L)$,
 $\Delta E_i(L)$ $[=E_i(L)-E_{\rm g}(L)]$,
 $\zeta$ $(=\sqrt3/2)$, and
 $f$ 
 correspond to the ground-state energy, an excitation gap, the geometric
 factor, and the free-energy density, respectively.
 $E_{\rm g}$ is found in the zero momentum sector ($k=0$).
 Also, $E_i$ is in a sector specified by its symmetry property.
 In the PRG calculation, we have employed the lowest excitation,
 corresponding to $\sigma$, with $k=2\pi/3$.
 We then numerically solve the condition
 $L\Delta E_\sigma(L)=L'\Delta E_\sigma(L')$
 with respect to $z$ for given values of $w$.
 We have extrapolated finite-size estimates of $L=6$, 9, and 12 to the
 thermodynamic limit based on an assumption of the leading $O(L^{-2})$
 correction
 \cite{Derr82}.
 Then, we find that the curve $z_2(w)$ continues smoothly to the open
 circle $w_{\rm BKT}$, which exhibits a consistency with the previous
 result
 \cite{Jaco09}.

 To check the universality class, we evaluate $c$ and $x_\sigma$ along
 $z_2(w)$.
 Plotted in Figs.\ \ref{CFT}(a) and \ref{CFT}(b) are the extrapolated
 results
 \cite{Qian05}.
 Although we cannot extract reliable data around $w_{\rm BKT}$ due to
 the smallness of the system size, we can find evidence to support the
 3-state Potts universality
 $c=\frac45$ and $x_\sigma=\frac{2}{15}$ down to $u\simeq -1.5$.
 Also, Fig.\ \ref{CFT}(c) provides a finite-size scaling plot of the gap
 $\Delta E_\sigma(L)=L^{-1}\Psi\{[z-z_2(w)]L^\frac{1}{\nu}\}$ at $u=-1$.
 Since $z-z_2$ linearly couples to $\varepsilon$, we expect the plot
 with $\nu=1/(2-x_\varepsilon)=\frac56$ to yield a collapse of the
 finite-size data onto a single curve.
 The result exhibits a good scaling property and thus supports our
 prediction.
 On one hand, for $u<-1.5$, we find deviations of data from the values,
 which implies that the system is approaching the tricritical point.
 We search it along $z_2(w)$, using the criterion
 $\myx{3}{3}{6}=\frac{2}{21}$ [see Fig.\ \ref{CFT}(b)]
 \cite{Qian05}
 and estimate $(w_{\rm t}, z_{\rm t})\simeq(\wTRI,\zTRI)$, which is
 given by the open square in Fig.\ \ref{phase}
 (the monomer density is estimated as around $\rmTRI$).
 To check its criticality, we estimate the central charge and scaling
 dimensions of the thermal operators. 
 The results are
 $c             \simeq \cTRI$,
 $\myx{1}{2}{6} \simeq \xonetwoTRI$, and
 $\myx{1}{3}{6} \simeq \xonethreeTRI$.
 We thus find reasonable agreement (within a few percent) between the
 numerical data and the theoretical predictions which strongly supports
 our RG argument.

 \ZUsan

 In the strong-doping region, we have found some data to imply a
 first-order phase transition, e.g., an abrupt change in the monomer
 density and a double-peak structure in the monomer-density distribution
 function.
 These could be used as means to determine the phase transition boundary
 $z_1(w)$.
 However, we provide here numerical solutions of the PRG equation,
 represented by filled squares with the curve in
 Fig.\ \ref{phase}
 because they are known to give accurate estimations
 \cite{Priv83Rikv82}.
 For large $z$ and $w$, we can analytically estimate the phase boundary
 via an energy comparison between the complete columnar state and the
 dimer vacuum.
 For the honeycomb lattice case, $z=w$, which is given by the dotted
 line in Fig.\ \ref{phase}.
 We observe that filled squares asymptotically converge to the line.

 In conclusion, motivated by recent experiments on IPM, we have
 investigated the MDM system on the honeycomb lattice.
 We have provided the global RG flow diagram, which determines
 universality classes of phase transitions.
 Also, we have performed numerical calculations and provided evidence
 that supports our predictions.
 Regarding the relationship to experiment, there exist some issues.
 For instance, the phase diagram of Dy$_2$Ti$_2$O$_7$ revealed a
 first-order phase transition between the kagome ice and the saturated
 state at $H\simeq0.9$ T for $T<0.36$ K
 \cite{Saka03}.
 In contrast, in our MDM system, the critical phase is not stabilized in
 the doped region.
 This implies that the long-range dipole interaction plays a crucial
 role for a stabilization of the kagome ice observed in real
 materials.
 However, this scenario is
 currently
 at the level of speculation, and
 its confirmation is left as a future work.

 The author thanks
 K. Goto,
 H. Kadowaki,
 G. Tatara,
 M. Fujimoto, 
 and 
 K. Nomura
 for stimulating discussions.
 Main computations were performed by using the facilities at the
 Cyberscience Center in Tohoku University.

 \newcommand{\AxS}[1]{#1,}
 \newcommand{\AxD}[2]{#1 and #2,}
 \newcommand{\AxT}[3]{#1, #2, and #3,}
 \newcommand{\AxQ}[4]{#1, #2, #3, and #4,}
 \newcommand{\AxP}[5]{#1, #2, #3, #4, and #5,}
 \newcommand{\AxH}[6]{#1, #2, #3, #4, #5, and #6,}
 \newcommand{\AxHp}[7]{#1, #2, #3, #4, #5, #6 and #7,}
 \newcommand{\AxO}[8]{#1, #2, #3, #4, #5, #6, #7 and #8,}
 \newcommand{\REF }[4]{#1 {\bf #2}, #3 (#4)}
 \newcommand{\JPSJ}[3]{\REF{J. Phys. Soc. Jpn.\           }{#1}{#2}{#3}}
 \newcommand{\PRL }[3]{\REF{Phys. Rev. Lett.\             }{#1}{#2}{#3}}
 \newcommand{\PRA }[3]{\REF{Phys. Rev.\                  A}{#1}{#2}{#3}}
 \newcommand{\PRB }[3]{\REF{Phys. Rev.\                  B}{#1}{#2}{#3}}
 \newcommand{\PRE }[3]{\REF{Phys. Rev.\                  E}{#1}{#2}{#3}}
 \newcommand{\NPB }[3]{\REF{Nucl. Phys.\                 B}{#1}{#2}{#3}}
 \newcommand{\JPA }[3]{\REF{J. Phys.\ A: Math. Gen.       }{#1}{#2}{#3}}
 \newcommand{\JPC }[3]{\REF{J. Phys.\ C: Solid State Phys.}{#1}{#2}{#3}}
 \newcommand{\IBID}[3]{\REF{{\it ibid.}}{#1}{#2}{#3}}
 \newcommand{\etal}{{\it et al.}}


\begin{thebibliography}{99} 
  \bibitem{Fowl37}
	  \AxD{R.H. Fowler}{G.S. Rushbrooke}
	  \REF{Trans. Faraday Soc.}{33}{1272}{1937}. 

  \bibitem{Kast61}
	  \AxS{P.W. Kasteleyn} 
	  \REF{Physica (Amsterdam)}{27}{1209}{1961}.

  \bibitem{Temp61Fish61}
	  \AxD{H.N.V. Temperley}{M.E. Fisher}
	  \REF{Phil. Mag.}{6}{1061}{1961};
	  \AxS{M.E. Fisher}
	  \REF{Phys. Rev.}{124}{1664}{1961}.

  \bibitem{Blun08}
	  \AxS{M.O. Blunt \etal}
	  \REF{Science}{322}{1077}{2008}.

  \bibitem{Jaco09}
	  \AxD{J.L. Jacobsen}{F. Alet}
	  \PRL{102}{145702}{2009}.

  \bibitem{Blot82}
	  \AxD{H.W.J. Bl\"ote}{H.J. Hilhorst}
	  \JPA{15}{L631}{1982};
	  \AxT{B. Nienhuis}{H.J. Hilhorst}{H.W.J. Bl\"ote}
	  \JPA{17}{3559}{1984}.

  \bibitem{Harr97Rami99}
	  \AxS{M.J. Harris \etal}
	  \PRL{79}{2554}{1997};
	  \AxS{A.P. Ramirez \etal}
	  \REF{Nature}{399}{333}{1999}.

  \bibitem{Hert00}
	  \AxD{B.C. den Hertog}{M.J.P. Gingras}
	  \PRL{84}{3430}{2000}.

  \bibitem{Mats02Hiro03}
	  \AxS{K. Matsuhira \etal}
	  \REF{J. Phys.: Condens. Matter}{14}{L559}{2002};
	  \AxS{Z. Hiroi \etal}
	  \JPSJ{72}{411}{2003}.
	  
  \bibitem{Saka03}
	  \AxS{T. Sakakibara \etal}
	  \PRL{90}{207205}{2003}.

  \bibitem{Moes01Udag02}
	  \AxD{R. Moessner}{S.L. Sondhi}
	  \PRB{63}{224401}{2001};
	  \AxT{M. Udagawa}{M. Ogata}{Z. Hiroi}
	  \JPSJ{71}{2365}{2002}.

  \bibitem{Cast08}
	  \AxT{C. Castelnovo}{R. Moessner}{S.L. Sondhi}
	  \REF{Nature}{451}{42}{2008}.

  \bibitem{Monopoles}
	  \AxS{S.T. Bramwell \etal}
	  \REF{Nature}{461}{956}{2009};
	  \AxS{D.J.P. Morris \etal}
	  \REF{Science}{326}{411}{2009};
	  \AxS{T. Fennell \etal}
	  \REF{Science}{326}{415}{2009}.

  \bibitem{Kado09}
	  \AxS{H. Kadowaki \etal}
	  \JPSJ{78}{103706}{2009}.

  \bibitem{Isak04}
	  \AxQ{S.V. Isakov}{K.S. Raman}{R. Moessner}{S.L. Sondhi}
	  \PRB{70}{104418}{2004}.

  \bibitem{Isak05Taba06}
	  \AxT{S.V. Isakov}{R. Moessner}{S.L. Sondhi}
	  \PRL{95}{217201}{2005};
	  for the kagome ice,
	  see also
	  \AxS{Y. Tabata \etal}
	  \IBID{97}{257205}{2006}.

  \bibitem{Alet05}
  	  \AxS{F. Alet \etal}
	  \PRL{94}{235702}{2005}.

  \bibitem{Papa07}
	  \AxT{S. Papanikolaou}{E. Luijten}{E. Fradkin}
	  \PRB{76}{134514}{2007}.

  \bibitem{Henl97}
	  \AxS{C.L. Henley}
	  \REF{J. Stat. Phys.}{89}{483}{1997}.

  \bibitem{Otsu09}
	  \AxS{H. Otsuka}
	  \PRE{80}{011140}{2009}.

  \bibitem{Lech02}
	  \AxT{P. Lecheminant}{A.O. Gogolin}{A.A. Nersesyan}
	  \NPB{639}{502}{2002}.

  \bibitem{Zamo87}
	  \AxS{A.B. Zamolodchikov}
	  \REF{Sov. J. Nucl. Phys.}{46}{1090}{1987}.

  \bibitem{Ludw87}
	  \AxD{A.W.W. Ludwig}{J.L. Cardy}
	  \NPB{285}{687}{1987}.

  \bibitem{Nien82}
	  \AxS{B. Nienhuis}
	  \JPA{15}{199}{1982}.

  \bibitem{Card84}
	  \AxS{J.L. Cardy}
	  \JPA{17}{L385}{1984}.

  \bibitem{Blot86}
	  \AxT{H.W.J. Bl\"ote}{J.L. Cardy}{M.P. Nightingale}
	  \PRL{56}{742}{1986};
	  \AxS{I. Affleck}
	  \IBID{56}{746}{1986}.

  \bibitem{Derr82}
	  \AxD{B. Derrida}{L.De. Seze}
	  \REF{J. Phys. (Paris)}{43}{475}{1982}.

  \bibitem{Qian05}
	  \AxT{X. Qian}{Y. Deng}{H.W.J. Bl\"ote}
	  \PRE{72}{056132}{2005}

  \bibitem{Priv83Rikv82}
	  \AxS{P.A. Rikvold \etal}
	  \PRB{28}{2686}{1983};
	  \AxD{V. Privman}{M.E. Fisher}
	  \REF{J. Stat. Phys.}{33}{385}{1983}.
 \end{thebibliography}
 \end{document}